\documentclass[pre,aps,superscriptaddress,twocolumn]{revtex4-1} 

\usepackage{amsmath}
\usepackage{graphicx}
\usepackage{subfigure}
\usepackage{soul}
\usepackage{bbold}
\usepackage{color}

\newcommand{\beq}{\begin{equation}}
\newcommand{\eeq}{\end{equation}}

\newcommand{\p}{\partial}
\graphicspath{{figures/}}

\begin{document}

\title{The equilibrium fluctuations of a semiflexible filament cross linked into a network}

\author{Jonathan Kernes} 
\affiliation{Department of Physics and Astronomy, UCLA, Los Angeles California 90095-1596, USA}

\author{Alex J. Levine}
\affiliation{Department of Physics and Astronomy, UCLA, Los Angeles California 90095-1596, USA}
\affiliation{Department of Chemistry and Biochemistry, UCLA, Los Angeles California 90095-1596, USA}
\affiliation{Department of Biomathematics, UCLA, Los Angeles California 90095-1596, USA}

\date{\today}

\begin{abstract}

We examine the equilibrium fluctuation spectrum of a constituent semiflexible filament segment in a network.  The effect of this cross linking is to modify 
the mechanical boundary conditions at the end of the filament.  We consider the effect of both tensile stress in the network, and its elastic compliance.  Most significantly, 
the network's compliance introduces a nonlinear term into the filament Hamiltonian even in the small-bending approximation.  We analyze the effect of this nonlinearity upon the 
filament's fluctuation profile.  We also find that there are three principal fluctuation regimes dominated by one of the following: (i) network tension, (ii) filament bending stiffness, or
(iii) network compliance. We propose that one can use observed filament fluctuations as a noninvasive probe of network tension, and we provide the necessary response function 
to quantitatively analyze this sort of ``tension microrheology'' in cross linked semiflexible filament networks.

\end{abstract}
\maketitle

\section{Introduction}
A variety of biological materials are composed of semiflexible filamentous networks, including F-actin, collagen, fibrin, 
and intermediate filaments~\cite{Pritchard2014,Broedersz2014,Chen2010}. Such networks have a rich linear rheology and 
exhibit a characteristic set of nonlinear mechanical features such as negative normal stress~\cite{Janmey2007, Kang2009}, nonaffine 
deformations~\cite{Heussinger2007, Fernandez2009}, and strain hardening~\cite{Storm2005, MacKintosh1995}. 
Because of these nonlinearities, tension propagation in filament networks appears to strongly deviate from the predictions of continuum elasticity theory, making it
difficult to predict both the interactions between molecular motors in cytoskeletal networks and 
between force-generating cells in the extracellular matrix.
\begin{figure}
\includegraphics[scale=0.6]{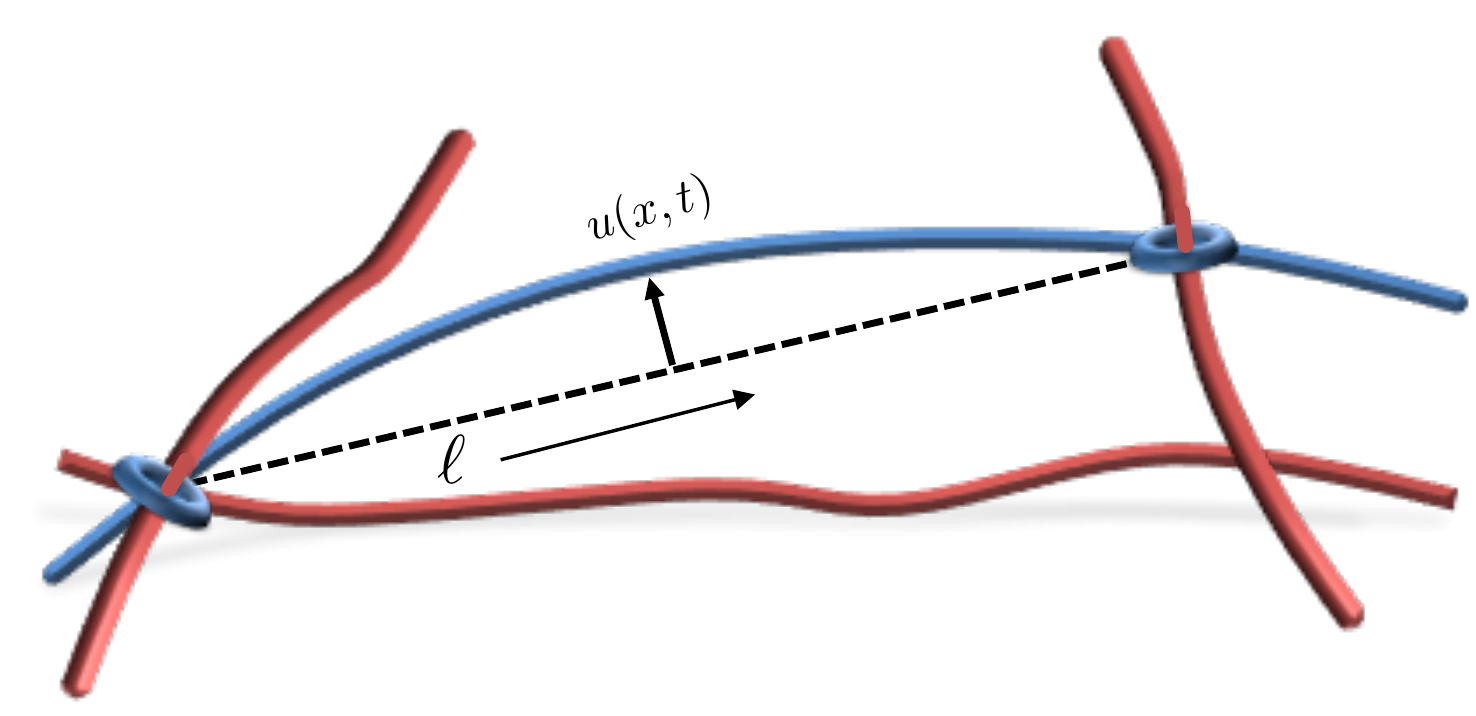}
\caption{(color online) A schematic diagram of a particular filament (blue) cross linked into a network of similar filaments (red). The cross links are represented by rings.}
\label{networkpic}
\end{figure}

Microrheology has been a useful probe the local mechanical properties of such systems and a number of techniques have 
been developed~\cite{Chen2010}. This approach uses the position fluctuations of 
tracer particles, measured using any one of a variety of instruments such as optical 
tweezers~\cite{Starrs2002,Addas2004,Latinovic2010} or laser deflection tracking~\cite{Gittes1997,Schnurr1997}, to extract the collective elastic response 
properties of the network.  A related microrheological approach that might allow one to map tension in filament networks, is to monitor 
the transverse undulations of network's filaments. Each filament's observed fluctuation profile is specified
by its intrinsic bending rigidity and applied tension so these measurements should produce a tension map in the network. To enable this sort of analysis, one 
must consider the predicted fluctuation spectrum of a filament cross linked into a network of similar filaments.  This cross linking to the network introduces 
new mechanical boundary conditions on the ends of the filament so that filament fluctuation spectrum not only reports on the filaments intrinsic mechanics, 
{\em e.g.}, bending modulus, but also on the collective mechanical compliance and stress state of the network to which it is coupled.  

In this manuscript we focus on this question of the role of the boundary conditions on filament fluctuations, showing that coupling the filament to an elastic network necessarily 
introduces a non-quadratic term in the filament's Hamiltonian, even in the small bending approximation.   The analysis of this issue, 
which is necessary to enable microrheological tension mapping in the filament networks, 
poses a few theoretically interesting problems explored here.

In the remainder of this manuscript, we explore the role of boundary conditions of various complexities, 
starting from the classic problem~\cite{MacKintosh1995} of a filament with its ends constrained to lie along one axis and subjected to a fixed tensile load.  Our
analysis culminates with the case in which the filament's end point is coupled to a combination of two Hookean (linear) elastic springs with differing 
spring constants such that one is oriented perpendicular and the other parallel to the 
undeformed filament's path.  This is the most general possible coupling of the filament to a linear elastic solid.  We do not consider the 
effect of applied constraint torques at the boundary, because we assume that the linker molecules are too small to provide significant torques. 
In addition, we allow a variation of the rest length of the longitudinal spring, enabling us to apply a fluctuating tension with non-vanishing mean to the system.  This 
allows us to explore how the local filament fluctuations report on the stress state of the network. 
We summarize our result as well as discuss experimental tests and dynamical extensions in section~\ref{sec:conclusion}.

\section{Semiflexible filament model} 
 \label{sec:model}

To study the effect of various boundary conditions on filament fluctuations, we will compute 
the two-point correlation function of the transverse displacement $u(x)$ of an element of a filament labeled by an arclength variable $x$. The
two-point function
\beq 
\label{eq:2point}
G(x,x') = \langle u(x) u(x') \rangle,
\eeq
is a natural extension of the particle mean-square displacement to filaments.  The 
angular brackets $\langle \ldots \rangle$ denote a thermal average. We do not here consider extensions of the analysis to nonequilibrium (e.g., motor driven
situations) but such extensions are, in principle, possible.

\begin{figure}
\includegraphics[scale=0.62]{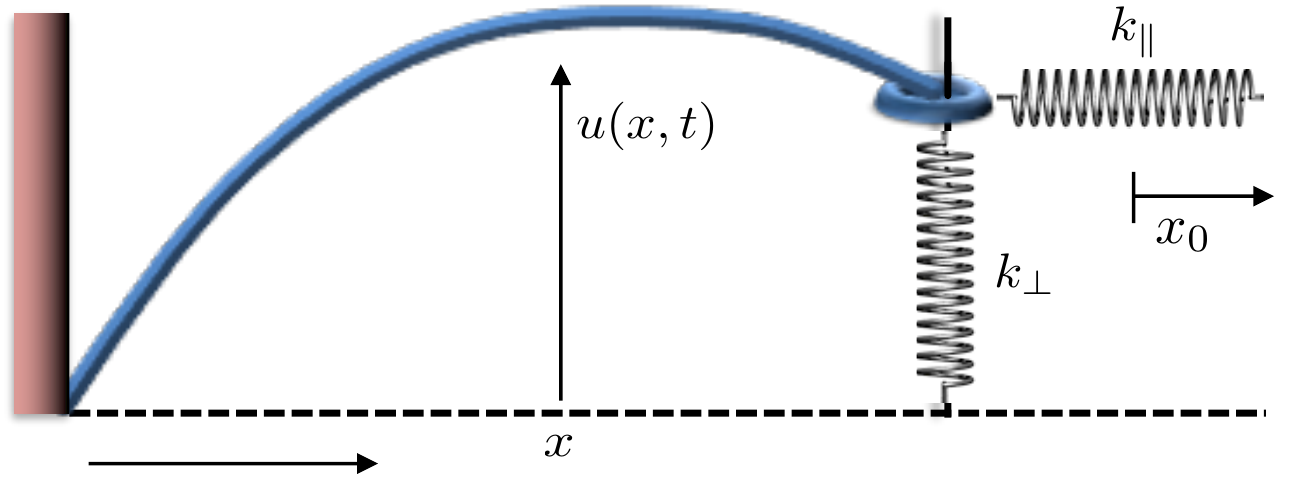}
\caption{(color online) Schematic diagram of a single semiflexible filament with the left endpoint pinned. Both endpoints are subject to torque-free boundary conditions.
The right endpoint is attached to a combination of a 
longitudinal spring of rest length $x_0$ with spring constant $k_\parallel$, and a transverse spring of zero rest length with spring constant $k_\perp$. These 
represent the elastic compliance of the network.}
\label{schematic}
\end{figure}

A schematic drawing of the filament is shown in Fig.~\ref{schematic}. We treat the filament as being  inextensible with contour length $\ell$ 
less than its persistence length $\ell_p \equiv \kappa/k_BT$. In this limit we may neglect states of the filament containing loops or overhangs  and 
describe its state of deformation by a two dimensional vector valued function $\bf{u(x)}$, giving the transverse displacement of  a 
material element of the filament parameterized by the arclength. To quadratic order in these displacements, the Hamiltonian admits two 
independent polarization states of these undulatory waves; we focus on just one of these here, replacing the vector $\bf{u(x)}$ by a 
scalar quantity  $u(x)$. In the presence of tension $\tau$, the elastic energy of deformation is given by~\cite{MacKintosh1995}
\beq \label{H0}
H_0 =\frac{1}{2} \int_0^{\ell} dx \left[\kappa u''(x)^2 + \tau u'(x)^2 \right] + \tau \ell.
\eeq
Here primes denote derivatives with respect to arclength $x$. To this order in $u$, we may neglect the distinction between that arclength and the 
projected length along the direction of the undeformed filament $\hat x$.  We here, and throughout this manuscript, take the 
range of integration to be over the projected length $\ell$ and hereafter suppress the limits of integration on such integrals.

Inextensibility demands that the contour length $\ell_\infty$ remain unchanged while filament undulations decrease the filament's projected length $\ell$ relative to the
contour length. Geometry relates the different between these two lengths, defining $\Delta \ell$ as 
\beq 
\label{DeltaL}
\Delta \ell \equiv \ell_{\infty} - \ell \approx \frac{1}{2}\int u'(x)^2 dx,
\eeq 
where we have kept terms in the integral up to ${\cal O}(u'^{2})$. 

The boundary conditions obeyed by the filament are found by applying the variational principle to Eq.~$\ref{H0}$. Due to the 
appearance of fourth order derivatives in the equation of motion, there are four equations to be satisfied. Pinning the left end 
of the filament at zero and demanding that the second derivatives $u''(x)$ vanish at both endpoints (the torque free condition) 
eliminates three of these. The remaining boundary condition corresponds to controlling either the transverse force $F^{ext}_\perp$ 
(conjugate to $u$)  or the  displacement of the right end point. The first choice of fixing the transverse force leads to
\beq
\label{transverse-force-bc}
-\kappa u'''(\ell) + \tau u'(\ell) = F_\perp^{ext},
\eeq
while pinning the transverse displacement at the right end leads to the simpler condition
\beq
\label{displacement-bc}
u(\ell) = 0.
\eeq
Using Eq.~\ref{transverse-force-bc} we may impose any number of forces on the end point that depend on that point's displacement. The 
most useful for our purposes is that of a Hookean spring (with zero rest length), which we implement by adding to the Hamiltonian
\beq
H_{k_\perp} = \frac{1}{2} ku(L)^2,
\eeq
which leads to the boundary condition 
\beq
\label{transverse-spring}
F^{ext}_\perp =-  k_\perp u(\ell).
\eeq 

We may also couple the right end of the filament to a longitudinal spring with rest length $x_{0}$, as shown in Fig.~\ref{schematic}. That 
spring puts the filament under a tension $\tau = k_\parallel (\Delta \ell+x_0)$, and contributes 
$H_{k_\parallel}=k_\parallel(\Delta \ell +x_0)(\Delta \ell +\ell)+\frac{1}{2} k_\parallel  (\Delta \ell+x_0)^2$ to the elastic energy of the 
combined filament and longitudinal spring -- see Eq.~\ref{H0}. 
The fact that the instantaneous tension acting on the filament depends on its deformation state introduces a nonlinear term into the energy
functional of the filament.  Because the tension depends on the difference between the projected length and arclength of the filament, which is 
given by an integral over the filament's configuration. The nonlinear term is also nonlocal.  Substituting the deformation-state dependent tension into 
the energy functional Eq.~\ref{H0}, we obtain 
\begin{eqnarray} 
\label{H}
H &=& \frac{1}{2} \int \left[\kappa u''(x)^2 + \tau u'(x)^2\right] dx +\\
 & +& \frac{k}{2}\int u'(x)^2 u'(y)^2 dx dy,
\nonumber
\end{eqnarray}
provided we identify
\begin{eqnarray} 
\tau &=& \tau_{\text{applied}} + k_\parallel (2 x_0 + \ell) \\ \label{tau}
k &=&3 k_\parallel/4,
\end{eqnarray}
where $\tau_{\text{applied}}$ accounts for any externally applied tension unrelated to the deformation of the longitudinal spring.   

In this 
calculation we assume that the change in tension along the filament is instantaneous.  This means that we treat the longitudinal speed of sound in the 
filament as being infinite, which is consistent with our inextensibility condition.  Presumably, this condition may be violated for very 
high wavenumber modes on very long filaments so that these modes relax faster than the tension propagation time.  

As mentioned above, there are other local but nonlinear terms associated with higher curvature configurations of 
the filament. We justify neglecting them by requiring the persistence length be sufficiently large.  The new nonlinear and nonlocal term introduced by the longitudinal spring
may not be neglected in this limit of stiff filaments. We emphasize that, while both the longitudinal and transverse springs affect the boundary conditions, 
only the longitudinal spring introduces nonlocal terms in the Hamiltonian.

Before studying the full problem, we first briefly review the properties of the 
equilibrium two-point function (Eq.~\ref{eq:2point}) for a filament with pinned transverse undulations at its endpoints~\cite{MacKintosh1995}. 
The term $\tau \ell$ in Eq.~\ref{H0} is a constant and may be ignored. The remaining pieces of the Hamiltonian are diagonalized by Fourier sine series
\beq
\label{sine-series}
u(x) \equiv \sum_p u_p \sin(p x).
\eeq
The zero displacement boundary condition -- see Eq.~\ref{displacement-bc} -- is satisfied by expanding the transverse displacements 
in half-integer wavelengths  $ p= \frac{n \pi}{\ell} \; n \in \mathbb{N}$.
The two-point function is then
\beq 
\label{G0}
G_{nm}^{(0)}= \frac{2k_B T/\ell}{\kappa p_{n}^4 + \tau p_{n}^2} \delta_{n m}.
\eeq

There is a cross over between curvature dominated modes at high $p$ and tension dominated ones for modes with 
wavenumber smaller than $\sqrt{\tau/\kappa}$.  Thus, tensed filaments admit a second length scale in addition to the thermal persistence 
length: 
\beq
\label{definition-tension-length}
\ell_t = \sqrt{\kappa/\tau}, 
\eeq
which we refer to as the {\em tension length}. An alternative description of this result is that the filament's fluctuations are governed 
primarily by bending provided that the tension is small in magnitude when compared to scale of the compressive force 
necessary to induce Euler-buckling: $\kappa/\ell^2 = k_{\rm B} T \ell_p/\ell^2$. 

Because of the Kronnecker delta linking the wavenumbers $p$ and $p'$ in the two-point function, it is straightforward to transform $G_{mn}^{(0)}$
back into position space to obtain $G_{0}(x,x')$. We find 
\beq
\label{G0-x-space}
G(x,x') = \frac{2k_{\rm B} T}{ \tau} \sum_{n=1}^{\infty}  \frac{\sin\left( n \pi x/\ell\right) \sin\left( n \pi x'/\ell\right)}{ \ell_{t}^{2}p_{n}^{4} + p_{n}^{2}}.
\eeq
We observe that the amplitude of mean square undulations $\sqrt{G(x,x)}$ peaks at the midpoint $\ell/2$, and that 
the fluctuation amplitude is dominated by the longest wavelength modes, which are on the order of the contour length $\ell$.

\section{Two-point function of filament attached to springs}
\label{sec:two-point-springs}
We now determine fluctuations of a filament attached to both a transverse and longitudinal spring, $k_\perp$ and $k_\parallel$ 
respectively, at its right end point.  These elastic couplings may be thought of as representing the elastic compliance of the network in which the filament is embedded. 
A sketch of such a situation is shown in Fig.~\ref{networkpic}. 
The schematic diagram corresponding to the single filament model is shown in Fig.~\ref{schematic}.
We begin by examining the effect of  each type of spring individually on the fluctuation spectrum of the filament, before 
considering their combined effect. 

\subsection{Transverse boundary spring}
We start with only a transverse spring. This spring shifts introduces a 
force-controlled boundary condition given by Eqs.~\ref{transverse-force-bc} 
and \ref{transverse-spring}.  The terms of the 
sine series introduced in Eq.~\ref{sine-series} no longer individually satisfy this boundary condition. This and the 
additional energy associated with the transverse spring constitute its full effect. 

We seek to compute the partition sum
\beq
\label{partition-sum-transverse}
Z = \int \mathcal{D}u e^{-\beta H}.
\eeq
Normally, this is accomplished by expanding the conformations of the filament in terms of the eigenfunctions of the Hamiltonian.  This expansion
makes the sum over states straightforward. The introduction of the more complex boundary condition at the right end of the filament makes
these eigenfunctions much more complicated than the simple sine series we used earlier.  In this subsection, we first show that one can still use 
the sine series and impose the transverse force boundary condition as a constraint on the infinite sum of the amplitudes of these sine modes.  
We then translate those constraints into a correction to the Hamiltonian, which now may be expanded in the sine series without further 
consideration of the problematic boundary condition. 

We begin by writing the partition sum, Eq.~\ref{partition-sum-transverse}, in terms of a sine series, and impose the boundary conditions by a delta function as 
\beq
\label{partition-sum-II}
Z = \int \prod_q d u_q \left[\delta\left(\sum_q F[u_{q}] \right) \right] e^{-\beta H[u_q]}.
\eeq
These boundary conditions introduce a constraint on the set of all the Fourier mode necessary to satisfy transverse force balance at the right hand 
side of the filament. We further assume that
the equation of constraint is a homogeneous function of the amplitude of the Fourier mode $u_{q}$ of first degree so that it may be written as
\beq
\label{psi-definition}
F[u_{n}] = \psi_{n}u_{n}.
\eeq
There is no sum over the repeated index. The form of $\psi_{n}$ depends on the boundary condition employed, but this form will always result as long as 
that boundary condition is a linear function of the displacement field and its derivatives.

We replace the delta functions by their limit as narrow Gaussians and thereby push the equations of constraint into the exponent, writing
\beq
\label{partition-sum-III}
Z = \lim_{\epsilon \to0} \frac{1}{\sqrt{4\pi \epsilon}} \int  \prod_n d u_n  \, e^{- \beta \sum_{n m} \left\{ \frac{ F[u_{n}] F[u_{m}]}{4\epsilon \beta } + H[u_{n}] \right\}}.
\eeq	
The boundary conditions now make up part of a new Hamiltonian of the filament $\tilde{H}[u_{q}]$ which is still quadratic in the 
$u$ fields but no longer diagonal in them. The effective Hamiltonian is given by 
\beq
\label{new-Hamiltonian}
\tilde{H}_{ nm} = \frac{1}{2} \left[G_{n m}^{(0)}\right]^{-1} u_{n}^{2} + \frac{1}{2} \left[G^{(1)}\right]^{-1}_{n m} u_{n}u_{m}.
\eeq
The purely diagonal part $G_{nm}^{(0)} \propto \delta_{nm}$ is given by Eq.~\ref{G0}.  The correction to this coming from enforcing the boundary conditions is 
\beq
\label{G(1)}
\left[G^{(1)}_{n m}\right]^{-1}=  \frac{\psi_n}{\sqrt{2 \epsilon \beta}}\frac{\psi_m}{\sqrt{2 \epsilon \beta}}.
\eeq
 For the case of a transverse spring attached to the right endpoint, we find that 
\beq
\label{psi-transverse-force}
\psi_n  = (-1)^n \left[\kappa p_n^3 + \tau p_n+ k_\perp \cos(n \pi)\sin(n \pi) \right].
\eeq
It appears that the last term in the above expression can be safely set to zero, but this amounts to an incorrect ordering of limits that will result in not 
enforcing the transverse force balance term correctly at the endpoint.  We return to this point below.

The Sherman-Morrison identity~\cite{Sherman1950} allows one to write the inverse of a matrix plus a dyadic as
\beq
(G^{-1} + v w^T)^{-1} = G - \frac{G v w^T G}{1+ w^T G v}.
\eeq
Using this, we invert the quantity $\left[ G^{(0)}\right]^{-1} + \left[ G^{(1)}\right]^{-1} $ and obtain
\beq
\label{corrected-two-point}
\tilde{G}_{nm}= G^{(0)}_{nm} - \frac{G^{(0)}_{nk}\psi_k \psi_l G^{(0)}_{lm}}{2\epsilon + \psi_k G^{(0)}_{kl}\psi_l}.
\eeq
This $\epsilon\to0$ limit is now well defined.  Taking that limit here, we arrive at the two-point function corrected for the transverse spring boundary conditions. 
\begin{figure} 
\includegraphics[scale=0.8]{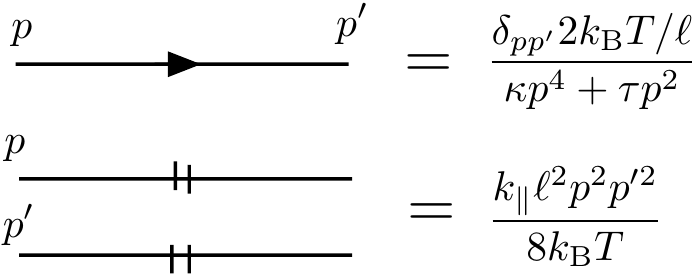}
\caption{The free propagator and nonlocal vertex. Slashes denote multiplication by momentum squared. 
The four field interaction depends on two independent momenta.}
\label{free diagrams}
\end{figure}

The result as written contains indeterminate parts equal to divergent sums multiplying zero.  We address these now.   We may write the two point 
function in Eq.~\ref{corrected-two-point} in form
\begin{subequations} 
\begin{align}
\label{SMformula1}
\tilde{G}_{nm} &= G^{(0)}(p_n) \delta_{nm}- \frac{F_n F_m}{D}  \\
D & = \psi_k G^{(0)}_{kl}\psi_l \\
F_n & = G^{(0)}_{nm} \psi_m.
\label{SMformula3}
\end{align}
\end{subequations}
The key term is $D$, which is given by
\beq
\begin{split}
D \beta&= \sum_{n=1}^{\infty} \left(\kappa p_n^2 + \tau \right) + k \ell \sum_{n=1}^{\infty} \frac{\sin 2 n \pi }{n \pi} + \\
&+ k^2\ell^2 \sum_{n=1}^{\infty} \frac{\sin^2 n \pi}{n^2 \pi^2}.
\end{split}
\eeq
D is clearly divergent. This arises because we are evaluating a Fourier-series outside of its radius of convergence, requiring us to analytically continue the sums. We 
start by noting that the third term is convergent so we may safely set it to zero using the sine function. The first term may be quickly calculated by noting 
continuations of the Riemann zeta function: $\zeta(0)=1/2$ and $\zeta(-2)=0$. Since the sums start at $n=1$ we find for the first line $-\tau/2$.  The 
second term is indeterminate. However, we notice that it is the Fourier sine series of the function $k\ell \left(\frac{1}{2}-\frac{x}{2L}\right)$, evaluated 
at $x=2L$. From this we observe that the sum must give $-k\ell/2$. 

Turning to the calculation of $F$ we find 
\beq
\label{F-definition}
\frac{2 \beta}{\ell} F_n = (-1)^n/p_n,
\eeq
as may be checked directly from Eq.~\ref{SMformula3} and the definitions of $G^{(0)}$ and $\psi$.
Putting these pieces together we find the corrected two-point function:
\beq 
\label{GBC}
\tilde G_{nm} = \frac{2k_B T/\ell}{\kappa p_n^4 + \tau p_n^2}\delta_{nm} + \frac{4k_BT/\ell}{\tau + k_\perp \ell}\frac{(-1)^{n+m}}{p_n p_m}.
\eeq
The two-point function decomposes into a sum of a diagonal part identical to that of the pinned filament -- see Eq.~\ref{G0} -- and an off-diagonal term, coupling
modes with different wavenumbers. This off-diagonal coupling results from the transverse spring boundary condition that introduces a coupling between 
various modes (labeled by wavenumber) since that boundary condition enforces a constraint on the sum of those modes. 

The off-diagonal term in Eq.~\ref{GBC} depends on the sum of two tensions: the externally imposed tension $\tau$ and a term proportional to the transverse 
spring constant $ k_\perp \ell$.  The magnitude of this term is controlled by the larger of these two tensions.  When both the tension and transverse spring constant 
both go to zero, we have the problem of a filament with a free end. The expansion of the system in terms of sines then fails, as is signaled by the divergence of
the two-point function.  We note that the real-space solution for the two-point function $G(x,x')$ for the case of a transverse spring can also be obtained, as 
shown in Appendix~\ref{app:realG}.

\subsection{Longitudinal spring}
We now consider a filament pinned at its right endpoint and attached to a longitudinal spring.  The Hamiltonian is Eq.~\ref{H}, with the boundary condition 
Eq.~\ref{displacement-bc}.   This time incorporating the boundary condition generates a nonlocal term in the Hamiltonian:
\beq 
\label{V}
V = \frac{k_{\parallel}}{2}\int u'(x)^2 u'(y)^2 dx dy,
\eeq
as seen in the second line of Eq.~\ref{H}. Despite this complication,  the two-point function calculation remains exactly solvable. 
We write the two-point function in terms of a perturbation theory in the parameter $k_{\parallel}$. Although the second term of Eq.~\ref{H} is not small, 
we will find that we can sum up all perturbative corrections to obtain a finite answer. 

The two-point function can be written as a sum over cumulants~\cite{Kardar2007}, 
\beq 
\label{cumulant}
\langle u(x) u(x') \rangle = \sum_{n=0}^\infty \frac{(-\beta)^n}{n!}\langle V^n u(x) u(x')\rangle_{0,c}.
\eeq
where  $\langle \ldots \rangle_{0,c}$ denotes the cumulant averaged with respect to the Hamiltonian in Eq.~\ref{H0}.  The 
perturbation series is most easily evaluated by reciprocal space.  We may organize the perturbation theory diagrammatically.  
The relevant diagrams are shown in Fig.~\ref{free diagrams}. The interaction vertex, Eq.~\ref{V}, is rather unusual.  It is
represented by a pair of disconnected propagators, 
with arbitrary wavenumbers $p,p'$ respectively.

\begin{figure} 
\includegraphics[scale=0.8]{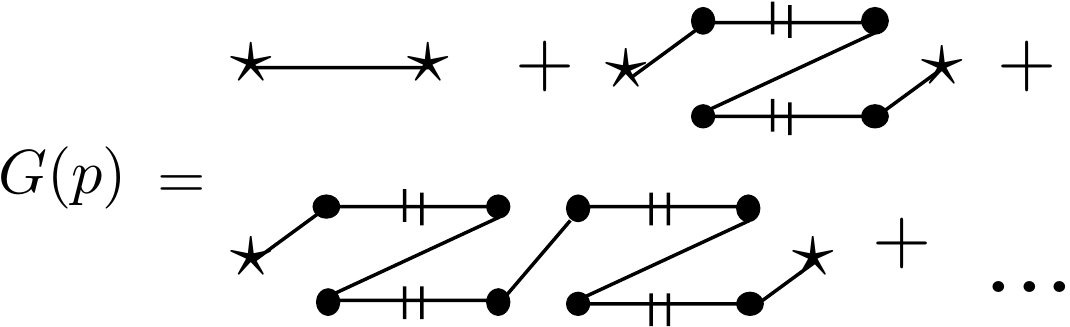}
\caption{The first three diagrams in the perturbative expansion of Green's function. No loops are possible in the connected diagrams, allowing the 
series to be resummed.}
\label{sample expansion}
\end{figure}

The series Eq.~\ref{cumulant} is shown diagrammatically for the first few terms in Fig.~\ref{sample expansion}.  
Consider the $m^{\rm th}$ order contribution to the two-point function. It is given by all possible contractions of $m$ vertices and two external legs. 
Due to the form of the interaction term, all loops are disconnected and thus do not contribute to the cumulant.  As a result, only lines contribute to the two-point function. 
Each diagram at $m^{\rm th}$ order is identical and 
equal to $\frac{\beta k_{\parallel} \ell^{2}}{8}p^4 G_0(p) G_0(p)^{2m}$,
with $G_0(p)$ defined in Eq.~\ref{G0}. The final step is to determine the 
combinatoric factor counting the number of identical diagrams at $m^{\rm th}$ order. Inspecting Fig.~\ref{sample expansion}, we find a 
total of $(4m)!!$ possible contractions at $m^{\rm th}$ order. Thus we obtain 
\beq
\label{up-correlation-sum}
\langle u_p u_{p'} \rangle = \delta_{pp'}G_0(p)\sum_{n=0}^\infty  \left(-\beta k_\parallel \ell^2p^4 G_0(p)^2\right)^n \frac{(4n)!!}{8^n n!}.
\eeq

The sum can be simplified by two identities. 
First  $(4n)!! 8^{-n} n!^{-1} = (2n-1)!!$. The second is $(2n-1)!! = (2\pi)^{-1/2}\int ds e^{-s^2/2} s^{2n}$. 
The second identity regulates the infinite sum in Eq.~\ref{up-correlation-sum}. Inserting these two identities and summing the resulting geometric series yields
\begin{equation} \label{intermediate}
G(p) = G_0(p) \sqrt{\frac{2}{\pi}}\int_0^\infty ds \frac{ e^{-s^2/2}}{1+\beta k_{\parallel} \ell^2 p^4 G_0^2 s^2},
\end{equation}
where we have written $G_{nm}(p) = G(p) \delta_{nm}$.
Performing this integral, we complete the calculation of the two-point function.  It is
\beq
\label{Glong}
G_{nm}(p)= \sqrt{\frac{\pi}{2\beta k_{\parallel} \ell^2}}\frac{e^{z_n^2}\text{Erfc}(z_n)}{p_n^2}\delta_{nm} ,
\eeq 
where we have introduced
\beq
\label{z-definition}
z_n = \frac{\kappa p_n^2 +\tau}{2\sqrt{2k_{\parallel}/\beta}}.
\eeq

\begin{figure}
\includegraphics[scale=0.55]{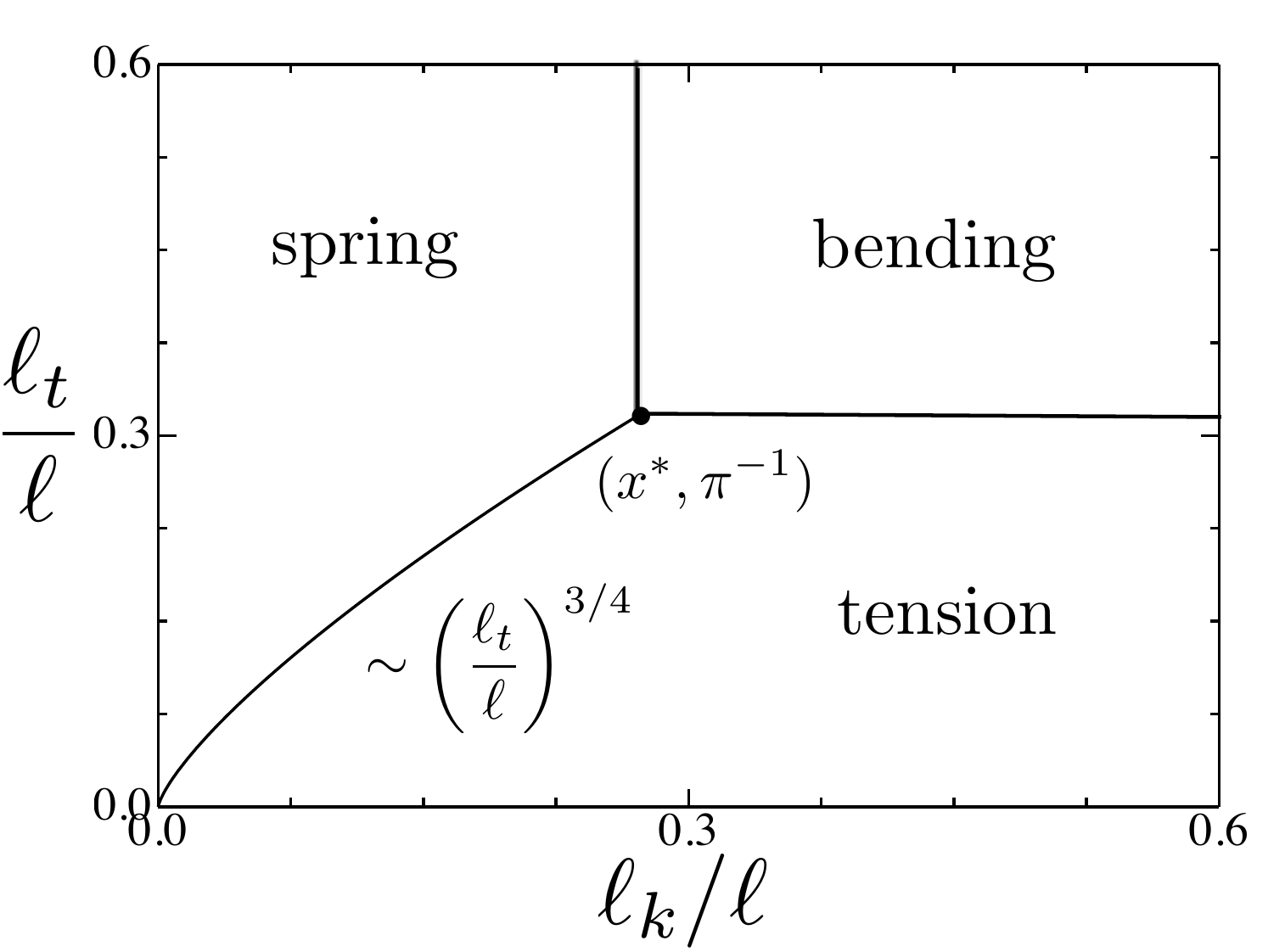}
\caption{Parameter space spanned by the ratios of the tension length and the longitudinal spring  length to the filament's length.  The three 
regions are defined by the type of boundary term that dominates the fluctuation profile: tension, network compliance (spring), or filament bending. 
The three regions coincide at the point $((2/z)^{1/3}\pi^{-4/3},\pi)$. The persistence length is set $\ell_p = \ell$ so
that $z=1$.}
 \label{fig:phases}
\end{figure}

In order to gain some physical insight into this result, we rewrite Eq.~\ref{Glong} as an integral by using the definition of the complimentary error function:
\beq
G(p)=\frac{1}{p^2}\int_0^\infty d\lambda e^{- \frac{1}{2} \beta \ell (\kappa p^2 + \tau)\lambda - \frac{1}{2}\beta k_\parallel \ell^2 \lambda^2}.
\eeq
The integral is dominated by its small $\lambda$ behavior.  Specifically, the integral is controlled by the value of $\lambda$ for which the argument of 
the exponential
\beq \label{arg}
- 2 \Phi(\lambda)  = \left(p^{2} \ell \ell_p  + \frac{\ell_p \ell}{\ell_t^2} \right) \lambda + \frac{\ell_p \ell^2}{\ell_k^3} \lambda^2
\eeq
equals one. 
Here we have defined a new length scale
\beq
\ell_k = (\kappa/k_{\parallel})^{1/3}
\eeq
governing the competition between bending and spring effects.  Considering the thermal persistence length and the 
tension length, the filament coupled to a longitudinal spring admits three independent length scales.
The dependence of the integral upon $p$ and these length scales is determined by which of the three terms in Eq.~\ref{arg}
reaches unity first with increasing $\lambda$.  There are clearly three possibilities generating three distinct results as shown in Fig.~\ref{fig:phases}.  

We fix the ratio of the persistence length to the total 
of the filament: $z = \ell_{p}/\ell$. Since the lowest order bending mode will dominate, we may replace the wavenumber $p \ell$ by $\pi$ in the following.  
In the spring dominated region the $\lambda^{2}$ term in $\Phi$ reaches unity before the other two 
terms (with increasing $\lambda$). This provides two inequalities. The first, it requires that $y= \ell_{t}/{\ell}$ is greater than $x^{3/4}  (z/2)^{1/4}$ where $x = \ell_{k}/\ell$.  
The second, it requires $x < x^{\star}= (2/z)^{1/3} \pi^{-4/3}$.  These provide the boundaries of the spring dominated regime (spring).  
Below and to the right of the spring dominated region lies
the tension-dominated regime (tension) in which the tension term $(z/2) \lambda y^{-2}$ reaches unity before the other two terms.  
This region extends to the right of $y = x^{3/4}  (z/2)^{1/4}$ 
and bounded above by $y = 1/\pi$.  Finally, the remaining part of the parameter space diagram is dominated by the longest wavelength 
bending mode.  This is the bending dominated regime (bending).  See Fig.~\ref{fig:phases}. 

The basic principle determining these regions is that the stiffest elastic element exerts the 
dominant influence upon the fluctuation amplitude. The corresponds to picking the shortest of the length scales associated with tension, bending, and 
network compliance introduced above.  The key signature of these three regimes can be understood as follows.   
Within the bending-dominated regime, the bending modulus dominates the 
amplitude of transverse undulations so that  $\langle u^{2} \rangle \sim \ell^3 T/\kappa$.  In the tension-dominated regime, the same undulations are 
controlled by the tensile stress in the network so we expect $\langle u^{2} \rangle \sim \ell T/\tau$.  Finally, in the region of 
parameter space where the network's compliance controls the amplitude of filament undulations, we expect to observe 
$\langle u^{2} \rangle \sim \ell \sqrt{T/k_\parallel}$, making the variance of $u$ in this regime proportional to $\sqrt{T}$.

In order to use the observed fluctuations for a filament-based tension probe, it is desirable to be in the tension dominated regime. 
For most semiflexible filaments of interest $z = \ell_{p}/\ell \le 1$.  As a result, the boundary 
$x^{\star}$ is typically quite small, resulting in a large tension-dominated regime.  Based on the boundary between the tension and network compliance (spring) 
dominated regions, we expect that the minimum observable tension should be $\sim \sqrt{T k_{\parallel}}$.  
In fact, the region of 
parameter space at small tension $y < \pi^{-1}$ where there is a transition from the tension-dominated 
fluctuation spectrum to the transverse spring dominated fluctuation spectrum (along the curve $y = \pi^{-1} (x/x^{\star})^{3/4}$) is likely to be hard to access 
experimentally.   All three regions, however, may be observable, particularly for sufficiently stiff filaments. 

\begin{figure}
\includegraphics[scale=0.65]{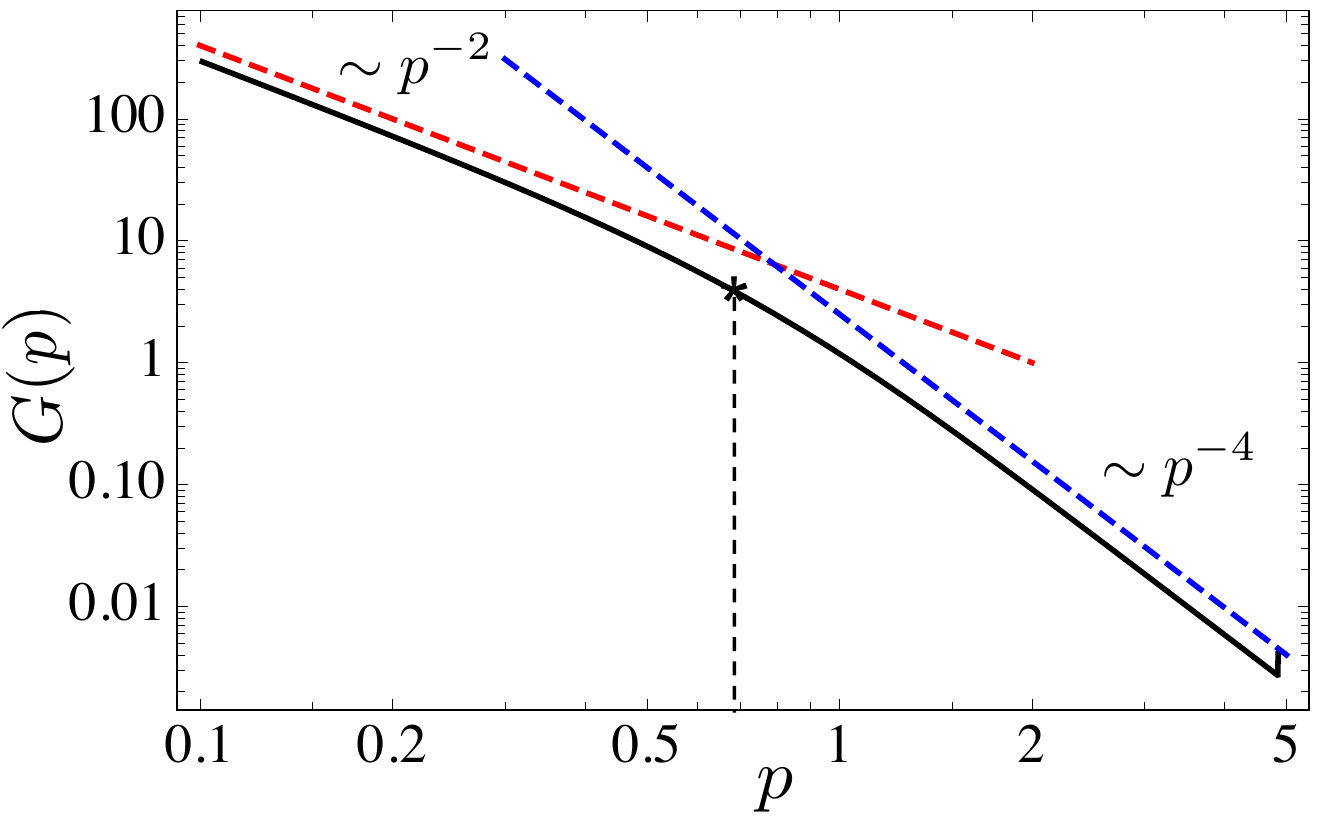}
\caption{(color online) Log-log plot of the two-point function $G(p)$ with respect to wavenumber $p$. There are two scaling regimes. Dashed lines illustrate slopes in the two regimes. At low 
wavenumber the function is dominated by a combination of tension and network
compliance, while at high wavenumber the function is controlled by the filament's bending stiffness.}
\label{loglogplot}
\end{figure}

For a fixed set of parameters we examine the scaling behavior of the two-point function with wavenumber $p$.  Using the result for the 
two-point function with a longitudinal spring in Eq.~\ref{Glong}, we make a log-log plot as shown in Fig.~\ref{loglogplot}.
For large $k_\parallel$, a series expansion shows that $G(p) \sim p^{-2}$, as expected for a tension-dominated filament. As for a simply pinned filament,
there is a transition with increasing wavenumber from this tension-dominated regime $G(p) \sim p^{-2}$, to a bending dominated one where $G(p) \sim p^{-4}$. 
The presence of the longitudinal spring changes the crossover point between these two regimes, when that spring constant is sufficiently large.  More precisely, 
the transition occurs at the usual wavenumber: $p^{\star} \sim \ell_{t}^{-1} = \sqrt{\tau/\kappa}$ when the longitudinal spring constant is less
than $k^\star =\frac{\tau^2}{2k_{\text{B}}T}$.   The transition from bending dominated to tension dominated modes should be experimentally accessible upon changing 
the longitudinal spring constant using a laser trap to hold one end of the filament.

In addition to the crossover between bending and tension dominated regimes, one may look for the mean tension in the filament.  This is perhaps the most 
\begin{figure}
\includegraphics[scale=0.57]{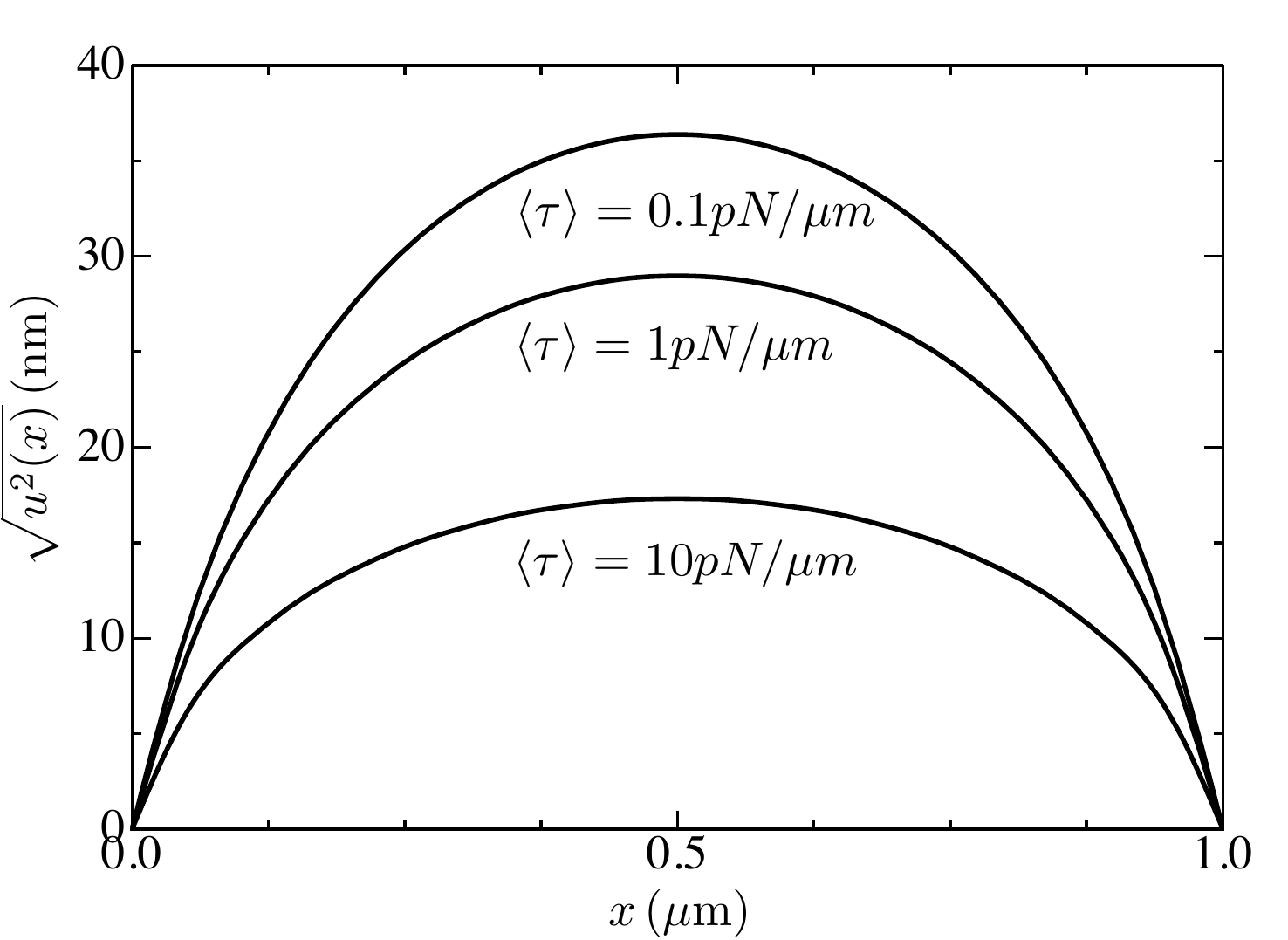}
\caption{Plots of the exact root mean square height-height fluctuations for parameters $\ell=1 \mu m, \ell_p=4/3 \ell, \ell_t = 0.1 \ell$. The 
average tensions of $0.1,\;1,$ and $10 (pN/\mu)$  correspond to lengths $\ell_k = 0.1123, \; .0498,$ and $.01937 (\mu m)$ respectively. 
The mean tension $\langle \tau \rangle$ is  calculated using $\langle \tau \rangle = k \langle \Delta \ell \rangle$.}
\label{fig:long-spring-u2}
\end{figure}
important theoretical result for the purposes of microrheological tension mapping in networks.  We can simply determine the mean filament tension from 
the relation 
\beq
\langle \tau \rangle = \tau_{\text{applied}} + k_\parallel \langle \Delta \ell \rangle.
\eeq
The average reduction of projected length of the filament due to thermal undulations can be directly computed from the 
two-point function via:  $\langle \Delta \ell \rangle = \frac{\ell}{2}\sum_p G(p) p^2$. 
We show in Fig.~\ref{fig:long-spring-u2} the expected fluctuation profiles for a range of values $\kappa$ and $\langle \tau \rangle$.  From these
one can compute the mean tension.

\section{Transverse and longitudinal springs}
Now we consider the combination of a longitudinal and a transverse spring attached to the right end point.   This represents the most complex boundary 
condition that will be encountered in a generic filament network. 
For this combination of springs, we have not found an exact solution, but we provide a self-consistent (Hartree) calculation for the fluctuation spectrum in 
which we replace the fluctuating tension in the filament by the longitudinal spring with its mean value obtained self-consistently in the calculation.

The essential step is to replace the quartic term in the Hamiltonian  Eq.~\ref{H} by
\beq
\label{eq: MFTdef1}
\frac{k_\parallel}{2}\left[\int u'(x)^2 dx\right] ^2 \to \sum_p k_\parallel \left[ \left(\frac{\ell}{2}\right)p^2 \langle \Delta \ell \rangle + \frac{1}{2}\left(\frac{\ell}{2}\right)^2 p^4 G_p \right] u_p^2.
\eeq
The second term results from the mixed term $\langle u'(x) u'(y) \rangle$.  We will show that it may be safely ignored.
Upon substituting Eq.~\ref{eq: MFTdef1}, the two-point function is immediately found to be
\beq
\label{eq:MFT 1}
\langle u_p u_p \rangle_{\text{MFT}}= \frac{2k_B T/\ell}{\kappa p^4 + \tau p^2 +2 k_\parallel \langle \Delta \ell\rangle p^2 + k_\parallel \ell p^4 \langle u_p u_p \rangle_{\text{MFT}}}.
\eeq

The final term in the denominator on the right hand side of Eq.~\ref{eq:MFT 1} depends on the full two-point function and must be satisfied self-consistently. 
In short, we replace:
\beq \label{MFTdef}
\frac{k_\parallel}{2}\left[\int u'(x)^2 dx\right] ^2 \xrightarrow[\text{MFT}]{} k_\parallel \langle \Delta \ell\rangle \int u'(x)^2 dx.
\eeq
Now we impose a self-consistency condition on the heretofore unknown value of $ \langle \Delta \ell\rangle$.  
This approximation is valid provided that the variance of $\Delta \ell$ is small, {\em i.e.}, $\sqrt{\langle \Delta \ell^2 \rangle_c} \ll \langle \Delta \ell \rangle $. 

The MFT Hamiltonian is of the form Eq.~\ref{H0}, but with  $\tau \longrightarrow \tau + k_\parallel \langle \Delta \ell \rangle$.  
The two-point function is found using our previous analysis of the transverse spring problem.  We write 
\beq 
\label{MFTG}
\langle u_p u_p \rangle_{\text{MFT}}= \frac{2k_B T/\ell}{\kappa p^4 + \tau p^2 +2 k_\parallel \langle \Delta \ell\rangle p^2}.
\eeq
We now impose the self-consistency condition by requiring that
\beq
\label{self-consistent}
\langle \Delta \ell\rangle = \frac{k_BT}{2}\sum_p \frac{1}{\kappa p^2 + \tau + 2 k_\parallel \langle \Delta \ell\rangle}.
\eeq

\begin{figure}
\includegraphics[scale=0.58]{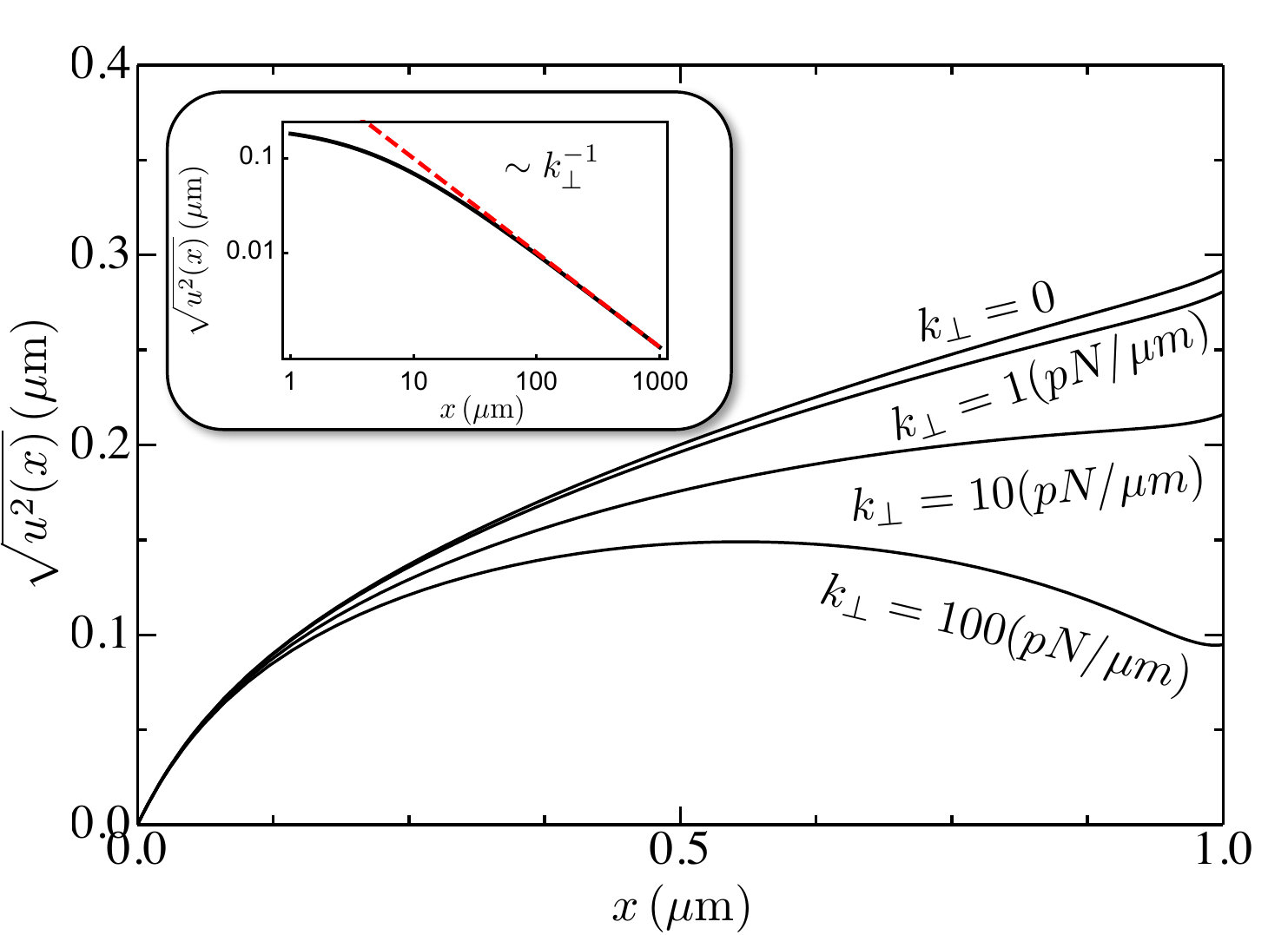}
\caption{(color online) The MFT root-mean square fluctuations for various values of the transverse spring constant $k_\perp$. 
The inset provides a log plot of the root-mean square fluctuations at spring (right hand side), as a function of $k_\perp$. 
For large large $k_\perp$, the endpoint fluctuation scales as $\sqrt{u^2(\ell)}\sim k_\perp^{-1}$, as expected for an ideal spring. 
Parameter values are $\kappa =0.0413 (pN \mu m), \; \tau =4.133 (pN), \;$and $k_\parallel=5 (pN/\mu m)$.}
\label{MFTforces}
\end{figure}

Because of the slow convergence of the sum, it is more convenient to solve the self-consistency condition Eq.~\ref{self-consistent} in position space. 
We note that Eq.~\ref{MFTG} is the Fourier-transformed Green's function associated with the equation of motion for $u(x)$, as can be obtained by the 
functional derivative of the self-consistent Hamiltonian.  This result, however, applies to the case in which we do not allow transverse displacements at the 
right end.  By changing this Green's function to the one appropriate for the transverse spring boundary condition while keeping the shift in tension, we can 
obtain the correct self-consistent condition for the case of a transverse spring (as well a longitudinal spring). The position space Green's functions for transverse boundary conditions, as well as their respective self-consistency conditions, are shown in Appendix~\ref{app:realG}. Particularly, we make use of Eqs.~\ref{eq: trans spring G} and~\ref{eq: trans spring lambda} to plot fluctuation dependence on transverse spring strength, as shown in Fig.~\ref{MFTforces}.

Fig.~\ref{MFTforces} plots the self-consistent two-point function $\langle u(x)^2 \rangle_{\rm SC}$ with a transverse and longitudinal spring.  We vary the 
spring constant of the transverse spring $k_\perp$ making several curves. Stiffer transverse springs clearly suppress endpoint fluctuations, but that 
suppression only decreases the variance of the endpoint logarithmically in $k_{\perp}$. Modest amplitude decreases require exponentially larger 
spring constants.

\section{Conclusion} \label{sec:conclusion}

We have analyzed the effect of various mechanical boundary conditions on the equilibrium fluctuation spectrum of a semiflexible filament.  This work expands upon the 
well-known case of  equilibrium filament undulations for a filament constrained to have its end points fixed to lie on an axis and with a prescribed tension. Specifically, we have considered the 
case in which there are Hookean elements constraining both the transverse and longitudinal displacement of an end in the presence of mean tension (and compression).   We have 
shown that one can directly account for the effect of a transverse harmonic pinning potential acting on the filament ends. More interestingly, the effect of the longitudinal harmonic potential acting on the 
projected length of the filament introduces a fluctuating tension, which is manifested as a nonlinear (quartic) term in the filament Hamiltonian, even in the case of filaments shorter than their 
own thermal persistence length, where geometric nonlinearities associated with the local filament curvature can be neglected. 
Understanding the implications of these boundary conditions for the observed fluctuations should enable one to quantitatively interpret the fluctuations of a filament segment cross linked to a 
network of such filaments in terms of the various model parameters, since the surrounding network acts to impose elastic constrains on the segment's end points.   

We point out that 
applying the elastic element to only one end should be irrelevant for this application to networks, since the energy of these ``network springs'' depends only on the filament's deviation from 
being straight and its projected length along its path in an unstressed state.  Moreover, the local effect of shear deformation should be to apply a local tension or compression. Even 
in the case of nonaffine deformation~\cite{Head2003,Frey2003}, where applied shear stress leads to local bending, we expect that the linearity of the 
response of the filament to bending (over small angles) will 
decouple of observed fluctuations from the mean bending. Thus, this analysis should allow one rather generally
to use the observed transverse fluctuations as sort of microrheological probe of tension propagation in networks, which we term ``tension microrheology.''

When we consider a filament subjected to a longitudinal elastic boundary condition that also imposes a finite mean tension, we note that there are three distinct regimes of fluctuations in which 
the scale of transverse undulations is controlled by one: (i) the elastic boundary condition, (ii) filament bending, or (iii) mean tension.  We have determined the boundaries of these parameter 
regimes, showing that tension dominates longitudinal compliance down to small tensions on the order of $\sqrt{T k_{\parallel}}$.  This result sets the minimum tension that may be resolved by 
the proposed ``tension microrheology.''  Using this result, we expect that for biopolymer networks with a modulus on the order of kPa and mesh size on the order of one micron, we will 
be able to resolve tensions down to $\sim1$pN.  This should enable the detection of both prestress in networks and molecular motor activity. For small affine network deformations, we may use the self-consistent longitudinal spring constant $k_\parallel$ to estimate the real part $G'$ of the network shear modulus. For a given cross section of the network, there are $\xi^{-2}$ segments, for $\xi$ the average network mesh size. This leads to a modulus $G \sim k_\parallel/\xi$~\cite{MacKintosh1995}.

The most direct experimental test of the theory is, however, in the analysis of the fluctuations of a single filament with one pinned end and Hookean constraints on the other.  This might be 
achieved using a filament bound to particles that are either optically or magnetically trapped. The trapping potential of the bead provides (approximate) Hookean boundary conditions, which are 
both adjustable and independently measurable.  As a result, the theory may be tested using a biopolymer filament of known bending modulus and measured length (e.g., F-actin) 
with no remaining fitting parameters.

Based on these calculations, one may imagine two directions for further study.  First, one may attempt a self-consistent evaluation of the compliance of the 
``network springs'' under the assumption that they represent a network of filaments identical to the one under consideration.  Such effective medium or 
mean-field theories have been pursued for networks of filaments and springs~\cite{Feng1985,das2007,Sheinman2012}.

Secondly, 
one may ask how the various boundary conditions affect the dynamics of filament undulations.  The presence of a longitudinal compliance once again renders 
the basic Langevin theory (with a local drag approximation~\cite{Hohenberg1977} or even slender body hydrodynamics~\cite{Lauga2009}) nonlinear.  
We intend to explore this question in future work in the limit of slow dynamics where the tension propagation time along the 
filament may be neglected.  Of course, this is consistent with our treatment here of the filament being inextensible.

\acknowledgements
The authors would like to thank the 
Botvinick group at UC Irvine for discussions leading to the idea of tension microrheology. 
We also thank Valentin Slepukhin for useful discussions.  This work was supported by NSF-DMR-1709785. 

\appendix

\section{Real space MFT Green's functions} 
\label{app:realG}
 The Green's function satisfies the equation of motion
 \beq \label{eom}
 \left[\kappa \p_x^4 - (\tau + 4 k_\parallel \langle \Delta \ell \rangle) \p_x^2 \right] G(x,x') = \delta(x-x')
 \eeq
To make the equations more readable, we define:
\beq
\lambda \equiv 2 \langle \Delta \ell \rangle.
\eeq
We solve for the Green's function by first finding solutions of Eq.~\ref{eom} in the regions $x\neq x'$. We then fix  the undetermined 
coefficients according to the prescribed boundary conditions and the jump discontinuity at $x=x'$ necessary to generate the delta function. 
The general solutions on the left (L) and right (R) of this discontinuity are
 \beq
 u_{L,R}(x) = A + Bx + C \cosh(p x) + D \sinh(p x)
 \eeq
 where 
\beq
p\equiv \sqrt{\tau+2k_\parallel \lambda /\kappa}
\eeq
and $\{A,B,C,D\}$ are, as yet, undetermined coefficients. 
We require that $u_L(x)$ and $u_R(x)$ be equal through the second derivative. The jump discontinuity then gives 
$u_R'''(x') - u_L'''(x') = 1/\kappa$.   Applying the boundary conditions at the discontinuity as well as at the prescribed boundaries 
yields an algebraic  system of equations from which the undetermined coefficients may be found. We obtain:
\begin{widetext}
\beq
G(x,x)_\text{pinned}=\frac{\sqrt{\kappa } \left(\coth \left(\sqrt{\frac{2 k_\parallel \lambda +\tau }{\kappa }}\right) \sinh ^2\left(x \sqrt{\frac{2 k_\parallel \lambda +\tau }{\kappa }}\right)-\frac{1}{2} \sinh \left(2 x \sqrt{\frac{2 k_\parallel \lambda +\tau }{\kappa }}\right)-(x-1) x \sqrt{\frac{2 k_\parallel \lambda +\tau }{\kappa }}\right)}{(2 k_\parallel \lambda +\tau )^{3/2}}
\eeq
\beq
G(x,x)_\text{free}=\frac{\sqrt{\kappa } \left(\coth \left(\sqrt{\frac{2 k_\parallel \lambda +\tau }{\kappa }}\right) \sinh ^2\left(x \sqrt{\frac{2 k_\parallel \lambda +\tau }{\kappa }}\right)-\frac{1}{2} \sinh \left(2 x \sqrt{\frac{2 k_\parallel \lambda +\tau }{\kappa }}\right)+x \sqrt{\frac{2 k_\parallel \lambda +\tau }{\kappa }}\right)}{(2 k_\parallel \lambda +\tau )^{3/2}}
\eeq
\beq
\label{eq: trans spring G}
G(x,x)_\text{$k_\perp$}= \left(\frac{x (k_\perp (-x)+k_\perp+2 k_\parallel \lambda +\tau )}{(2 k_\parallel \lambda +\tau ) (k_\perp+2 k_\parallel \lambda +\tau )}-\frac{\sqrt{\kappa } \text{csch}\left(\sqrt{\frac{2 k_\parallel \lambda +\tau }{\kappa }}\right) \left(\cosh \left(\sqrt{\frac{2 k_\parallel \lambda +\tau }{\kappa }}\right)-\cosh \left((1-2 x) \sqrt{\frac{2 k_\parallel \lambda +\tau }{\kappa }}\right)\right)}{2 (2 k_\parallel \lambda +\tau )^{3/2}}\right).
\eeq
\end{widetext}

To write the final answer, we must determine $\lambda$. The self-consistency condition is 
\beq
\lambda = \int_0^L dx \lim_{x'\to x} \p_x \p_{x'}G(x,x'). 
\eeq
For each of the three cases we find
\begin{widetext}
\beq
\lambda_{\text{free}}=\frac{\kappa  \sqrt{\frac{2 k_\parallel \lambda +\tau }{\kappa }}+(2 k_\parallel \lambda +\tau ) \coth \left(\sqrt{\frac{2 k_\parallel \lambda +\tau }{\kappa }}\right)}{2 \sqrt{\kappa } (2 k_\parallel \lambda +\tau )^{3/2}}
\eeq

\beq
\lambda_{\text{pinned}}=\frac{(2 k_\parallel \lambda +\tau ) \coth \left(\sqrt{\frac{2 k_\parallel \lambda +\tau }{\kappa }}\right)-\kappa  \sqrt{\frac{2 k_\parallel \lambda +\tau }{\kappa }}}{2 \sqrt{\kappa } (2 k_\parallel \lambda +\tau )^{3/2}}
\eeq

\beq
\label{eq: trans spring lambda}
\lambda_{\text{spring}}=\frac{\kappa  \sqrt{\frac{2 k_\parallel \lambda +\tau }{\kappa }}+(2 k_\parallel \lambda +\tau ) 
\coth \left(\sqrt{\frac{2 k_\parallel \lambda +\tau }{\kappa }}\right)}{2 \sqrt{\kappa } (2 k_\parallel \lambda +\tau )^{3/2}}-\frac{k_\perp}{(2 k_\parallel \lambda +\tau ) (k_\perp+2 k_\parallel \lambda +\tau )}.
\eeq

\end{widetext}

\bibliography{references2}

\end{document}